\documentclass[aps,prl,preprint,superscriptaddress]{revtex4}

\begin{document}






%\preprint{}


%\title{COMETS IN INDIAN SCRIPTURES}
\title{COMETS IN ANCIENT INDIA}










\author{Patrick Das Gupta}

\email[]{patrick@srb.org.in}


%\thanks{}


\affiliation{Department of Physics and Astrophysics, University of Delhi, Delhi - 110 007 (India)}







%\date{\today}

\begin{abstract}

The Indo-aryans of ancient India  observed  stars and constellations for ascertaining auspicious times in order to conduct  sacrificial rites ordained by  vedas. It is but natural that they would have recounted in the vedic texts about comets. In Rigveda ($\sim $ 1700 - 1500 BC) and Atharvaveda ($\sim $ 1150 BC), there are references to dhumaketus and ketus, which stand for comets in Sanskrit.  Rigveda mentions a fig tree with roots held up in the sky (Parpola 2009, 2010). Could it have been inspired by the hirsute appearance of a comet's tail?  Similarly, could `Ketu' (the torso or the  tail part of Rahu) be a Dravidian loan word, since `kottu', an old Tamil word, is associated with  scorpion's sting and top tuft of hair?  Varahamihira in 550 AD and Ballal Sena ($\sim $ 1100 - 1200 AD) have described a large number of comets recorded by ancient seers such as Parashara, Vriddha Garga, Narada, Garga, etc. In this article, I conjecture that an episode narrated in  Mahabharata of a radiant king, Nahusha,  ruling the heavens, and later turning into a serpent after he had kicked the seer Agastya (also the star Canopus), is a mythological retelling of a cometary event.
\end{abstract}


\pacs{}


%\keywords{}


\maketitle



\section{1 Introduction}
 
Barring the regular waxing and waning of the Moon, ancient
 observers seldom witnessed   celestial objects undergoing
 metamorphosis. In the pre-telescope era, our ancestors  were
 treated to such rare spectacles only on two occasions,  during
 (a) the solar/lunar eclipses   and  (b) cometary sightings wherein a gradual growth of a tail is seen, as the   comet approaches  Sun.

 Eclipses, of course, have been referred to in many ancient texts (see, for example, Stephenson 1997). 
Darkness inflicted on Sun and Moon   caused confusion and fear among people, as can be inferred from   Rigvedic allusions to  a solar eclipse (Kochhar 2010).
A proper understanding and  reasonably accurate predictions
of  these periodic phenomena had to wait till the arrival of the Greek astronomer Hipparchus (190 BC)  and, in India, of Aryabhatta (476 AD) on the scene (see, for example, Ansari 1977).
On the other hand, cometary appearances are  sporadic, and relatively rare due to their highly  eccentric orbits around the Sun,  with very large semi major axes.   A  bright comet with a   tail that grows and arches as it nears the Sun must have been  captivating and mysterious to  our forefathers.

 Comets are made of  dust, organic compounds like frozen methane and ammonia, ice as well as dry ice. They have  rocky nuclei of size $\sim $ 1- 10 km weighing $\sim 10^{15} - 10^{18} $ kg. Because of  small mass, comets have very low surface gravity so that the pressure exerted by the solar wind plasma pushes  the ionized and volatile  substances  outwards as they come within few AUs  (1 AU  $\simeq 1.5 \times 10^8 $ km) of Sun, leading  to tail formation.
 Recently, the tail  of the comet
 Siding Spring C/2013 A1 had a brief encounter with Mars during  October 19-20, 2014, that was monitored
 {\it in situ} by both NASA 's fleet of martian spacecrafts (that
 included MAVEN)  as well as Indian Mars Orbiter Mission.
 At the time of writing this article, European Space Agency's   Rosetta spacecraft, that has been cruising in space for about a decade,  succeeded  in lowering its lander Philae   on  the $\sim $ 4 km wide rocky nucleus of comet 67P, on November 12, 2014.
 
Many comets of antiquity have  been described in the astronomical records of the Greeks and the Chinese (Fotheringham 1919). In the Indian context,  it is plausible  that comets have been referred to  in  Rigveda, since  the Sanskrit word dhumaketu (meaning literally `smoke banner') appearing  in    about  half a dozen hymns, eventually came to mean  comet in Indian  languages. In the subsequent vedic literature, comets  have been highlighted,  with various terms  like Ketu,  Dhumaketu, Shikhi or Keshi  employed to describe them (Chandel \& Sharma  1991, Iyengar 2006, 2010).

 \section{2 Vedic literature  and Comets}

Vedas - Rig, Yajur, Saama and Atharva, composed in archaic  Sanskrit, are the oldest deciphered Indian texts. The word veda in Sanskrit means (sacred) knowledge, with cognates in other  Indo-European languages, such as    (w)oida in Greek,  wit (witness) in English and wissen in German (Witzel 2003). Rigveda ($ \sim $ 1700 - 1500 BC), the oldest of all,  consists  of about 1028 hymns invoking mostly nature gods,  goddesses and sacrificial rites. It also includes secular observations, vignettes of Indo-aryan life,  their desires, struggles and battles (Griffith 1896). From the first mention of iron in Atharvaveda,  the mantra portions of this text have been dated to about 1150 BC (Witzel 1995)

Some written documents of the Mitanni dynasty rule (1500 - 1350 BC) that contain  Rigvedic  terms were found in northern Syria. A person named  Kikkuli of this period had authored a horse training manual involving numerous  archaic Sanskrit terms. More surprisingly, a treaty between a Mitanni king and  a Hittite monarch around 1380 BC invoked major Rigvedic gods like  Indra, Varuna, Nasatya and Mitra as witnesses for the accord (Anthony 2007).  From the Mitanni agreement, antiquity of  Rigveda is confirmed (Witzel 1989, Kochhar 2000)    

Archaeoastronomical arguments indicate that vedas  have preserved even older traditions.
  In vedic texts, nakshatras or `lunar mansion'
(bright stars and constellations lying along Moon's path) have an exalted status (Subbarayappa 2008).   Krittika (Pleiades) is often listed as the first asterism among the 27 or 28 nakshatras,  which  according to Jacobi (1909), was due to its rising on the east during the vernal equinox, when  these rituals were getting established.  Because of  the precession of equinoxes, it is Pisces  at present, and not Pleiades, that rises on the east during the Spring equinox. This strongly suggests   that the  vedic practices  began in the period 3000 - 2000 BC (Jacobi 1909). 

Shatapatha Brahmana (SBr), $\sim $ 1000-700 BC, instructs one to light  an  auspicious sacrificial fire at the time of  Krittika's  rise  in the east  since Pleiades `rises invariably in the due east'  (Chattopadhyaya 2008). SBr was obviously  restating an ancient  observation that had no bearing on the contemporary location of Krittika during its  composition.  Even Chinese annals of 2357 BC  had recorded   that Alcyone, the bright central star of  Pleiades, was near the vernal equinox (Allen 1963, Chattopadhyaya 2008). 
Moreover, older vedic literature referred to  the  Big Bear constellation  as rikshas (bears, in archaic Sanskrit)   indicating a common Indo-European origin of the vedic people. By about 900 BC,  the stars of Ursa Major had been identified with seven Rigvedic seers, and hence  the name  Saptarishi or seven sages   (Ghurye 1972).

 Rigveda  also narrates Sun getting pierced thoroughly with darkness by a demon's son,  named Svarbhanu,  till Atri restores Sun's brilliance through sacred chants and offerings (Griffith 1896).      
 The  Atris were a powerful priestly clan, who had made substantial contributions to the corpus of vedic hymns  (Kochhar 2010). The  Svarbhanu episode portrays the occurrence of an ancient solar eclipse (Sengupta 1947, Markel 1990, Kochhar 2000, Yano 2003, Kochhar 2010, Vahia \& Subbarayappa 2011).

Vedic rituals  required strict temporal order.  Jyotishavedanga (also referred to as Vedangajyotisha),  ascribed to the commentator Lagadha, described the movements
of Sun and  Moon with respect to the nakshatras (lunar mansion) and stated categorically that positional astronomy's sole purpose is to determine the times for  performing vedic rituals (Pingree 1981, Yano 2003). Although  Pingree (1981) had dated Lagadha to about 400 BC, some authors  have placed the commentator   earlier than 1100 BC (Achar 2000).

Apart from innate curiosity and determination of time for sacrificial rites,  other reasons for keenly observing the night sky could have been the natural tendency to  be drawn to 
lustre, radiance,  golden  and silver rays from Sun and Moon, etc. on one hand, and to eschew  darkness on the other. This is amply evident from  Rigvedic verses, as seen from the numerous
hymns in praise of Agni (fire), Surya (Sun), Ushas (the dawns)
and the Asvin twins or Nasatyas ($\beta $ and $\gamma $  Arietis).  The mortal enemies of the Indo-aryans -  Dasas and Dasyus, were
portrayed as dark people.
Rigveda praises  Asvins ($\beta $ and $\gamma $  Arietis),  the
earliest deliverers of light,  for saving some of the seers like 
Atri, Kanva, Rebha, Vandana,   and Antaka from deep pits at different times, as well as rescuing Tugra's son Bhujjyu  from drowning in the sea with a ship's aid.
These lores signified possibly the role played by the twin stars in
giving a sense of direction and hope to the despondents,  before the breaking
of dawn.

The above discussion makes it  clear that for the  seers of antiquity, the night sky and its attendants  were  sacred. It is highly unlikely that comets of the past would have gone unnoticed. Asko Parpola (2009) has drawn attention to a  late Rigvedic verse that speaks of  an Indian  fig tree whose aerial roots  are held up in the sky by the god Varuna (guardian of cosmic law). In the later Puranic texts (containing old Hindu royal genealogies and mythologies), the phenomena of gravity-defying stars and planets  going round the fixed Dhruva (Pole star) is  explained by claiming that  these celestial objects are fastened to  invisible  rope like aerial roots growing outwards from the north star (Parpola 2009, 2010). This immediately raises the  question - was this  `aerial roots' simile inspired from the  hairy tail of a   comet of bygone era?

One would  expect  that the `lustre eulogizing' Indo-aryans  would  be enamoured of comets. On the contrary, comets and meteors were  often thought to be associated with impending doom as inferred from   Mahabharata and Puranas   as well as from Varahamihira's  Brihat Samhita (Bhat 1981).
A natural explanation is that appearances of comets and meteors
 are unpredictable. In the words of the renowned Indian
 astrophysicist M. N. Saha: `Comets appear from nowhere and on
 account of their weird appearance had always been  taken to
 portend great calamity' (Saha 1953). 
Meteors at times do lead to conflagrations and casualties. It is possible that the comets with their appendages did get associated with meteors, even in the past, for shooting stars do appear to have tails.  Ketus (in plural) have also been discussed in Atharvaveda denoting rays of light or fire-smoke combine, and could have represented comets or meteors (Kochhar 2010).  A verse from  Atharvaveda that links  dhumaketu (smoke banner)  to  mrityu (death), presumably because of its resemblance to the rising smoke   from a funeral pyre (Whitney 1905, Iyengar 2006)  may be hinting at yet another wrong  reason for comet-phobia among our ancestors.

R. N. Iyengar (2010), observing  that  the word dhumaketu appears seven times (mostly associated with maruts) in Rigveda, and that   Atharvaveda contains  a hymn about  Saptarishi  (Ursa Major) being veiled by a dhumaketu,  has concluded that dhumaketus referred to comets even in vedas. According to him, dhumaketus and maruts  represent  comets and meteor showers, respectively  (Iyengar 2010).  
However, the standard and more natural  interpretation of maruts is that they  were minor gods   associated with storms and cyclones who, at times, unleashed calamities on mortals. On the other hand, Indra in Rigveda was by far the most powerful nature god (of torrential rain, thunder and flood) and was likened to a solitary  bull-like warrior,  as can be reckoned from a hymn in which  the maruts implore the seer Agastya to intercede on their behalf and  request  Indra not to  slay them  (Griffith 1896).  This Rigvedic  hymn has also been interpreted differently  where it has been argued that it alludes to  seer Agastya's  bringing about a reconciliation between two warring tribes - Indo-aryans (worshippers of  Indra) and non-aryans (who prayed to  maruts)  (Ghurye 1977, Mahadevan  1986). 

Signs and symbols used by the pre-vedic  Indus Valley people  ($\sim $ 2500-1900 BC) of Harappa,  Mohenjo-daro, and of scattered north-western regions of India,  have still remained undeciphered. These symbols are found predominantly  on the steatite Indus seals. Parpola, while discussing the  seal M-414, has made a very interesting conjecture that its first sign could represent the sting of a scorpion and   therefore, the associated word could be the Proto-Dravidian word `kottu' meaning `sting' or `stinging' or  similar sounding words, which mean `top tuft of hair', `crest of a bird', or `pointed tip', etc. (Parpola 2009).  One could then ask: is the Sanskrit word  `ketu' a Dravidian loan word related to `kottu'? After all, `shikhi' another vedic word for comet, also means `with tuft of hair'. Moreover, comets have tails and the sting of a scorpion ensues from the latter's tail.

The epics too refer to comets. In Ramayana, deadly missiles are likened to comets. For instance, Ramayana mentions that when Dasaratha (father of Rama) and Kaikeyi (Rama's stepmother) were fighting the demons, Dasaratha was gravely  injured by  a comet like missile. In another place, Ravana hurls a missile that resembles a small sun like comet that  fatally wounds Lakshmana (Rama's brother). Similarly, in Mahabharata, Veda Vyasa (the author of the epic as well as an editor  of Rigveda) warns the blind king, Dhrtarashtra, about the  ill-fated Pandava-Kaurava war, citing ominous signs like the nakshatra Pushya
 ($\gamma $, $\delta $ and $\theta $ Cancri) being obscured by the Dhumaketu (Kochhar 2010).

\section{3 Of  Rahu, Ketu and Ketus}

The name of the eclipse causing demon, Svarbhanu,  of Rigveda  got transformed, over time,   to  Rahu.     Actually as Sun's foe, Rahu had made his debut earlier in Atharvaveda (Kochar 2010). This possibly led to a mix up between Svarbhanu and Rahu, as vedas were traditionally passed on orally. Elaborate myths concerning Rahu were narrated   in  later texts like Mahabharata, Bhagavata Purana and Vishnu Purana. Rahu had a serpentine form  and,  in a clandestine manner,  it had partaken 
Amrita, a celestial  ambrosia (making  the gods immortal) that had emerged out of the churning of Ocean. However, the Sun and the Moon gods had witnessed Rahu's deed. So, the demon tried  devouring  them, whereupon 
Vishnu  severed  its head  by hurling his deadly discus, the Sudarshan chakra, at Rahu.

  The head retained the name Rahu whereas  the torso with a tail was christened Ketu. Now, the comets were already being referred to as ketus in the vedic literature. The obvious reason  for naming  the torso  Ketu   is that both have tails. 
As the demon had already consumed some amount of  the ambrosia, Rahu as well as  Ketu had become  immortal. In the absence of a  torso, the Sun or  the Moon could not be retained for long after being  swallowed by the head Rahu. Eventually  they had to emerged out. That was the way  Hindu mythology dealt with  the phenomena of eclipses. 

Since Rahu tried to  seize  Moon and Sun, it got listed as a graha (leading to  seizure or grabbing, but the term later became synonymous with planet). Even the Sun and the Moon came to be known as grahas in Puranic and astrological texts. 
Gargyajyotisha, composed between BC and AD, probably by a descendant of  the seer Garga,  includes both 
 Rahu and Ketu in its list of nine planets, with Ketu representing comets and
not the severed torso of Rahu (Yano 2003). Was this because Garga, $\sim $ 100 BC (Kane 1975),   had a penchant for  observing comets, and had made a list of 77 comets that were characterized by a dark reddish hue, as mentioned in Varahamihira's Brihat Samhita (BS) of 550 AD?   Brihajjataka by Varahamihira  lists Rahu and Ketu as planets, with Shikhi as another word  for Ketu (Kochar 2010). 

Atharvaveda-Parishishtha, with many of its chapters composed after Greek astrology was introduced in India around 300 AD,  contain verses  not only  about grahas, nakshatras, rahu but also about  ketus  (comets) classified  according to seasons (Miki and Yano 2010). Generally, Hindu temples have sculptures representing  nava grahas (nine `planets') with Ketu depicted as having an anthropomorphic bust with  a tail.  Did Ketu  become a graha  because of its possible association with the old Tamil word `kottu' (scorpion's sting) that has been discussed in section 2? After all, scorpions too, like the crabs,  possess pincers to grab.  Parpola (2009) has conjectured that the crab sign found in many Indus seals depict grahas (those who seize).

Aryabhatta in 499 AD, dispensing  with Rahu and Ketu, gave the correct reason for the eclipses.  The plane of Moon's orbit (around  Earth) is inclined with respect to the Earth-Sun orbital plane leading to a line of intersection. Total eclipses happen only when  Moon and Sun get to be on this line at the same time, i.e. only when the Moon is either at the ascending or the descending node. Varahamihira  provided a clearer explanation for the eclipses, but called the lunar nodes Rahu and Ketu. A great deal of mathematical techniques pertaining to trigonometry flourished in many parts of the world as a result of human preoccupation with eclipses. So, one wonders whether the chance coincidence of angular diameters of Sun and Moon being almost the same in contemporary times (leading to the total eclipses) played a significant evolutionary role in  advancing  mathematics.  

As regards cometary records, BS  had cited (besides Garga)  the texts due to ancient seers like Parashara,  Vriddha Garga  and Narada. But their compositions on comets are no longer extant and, therefore, one has to fall back on the writings of Varahamihira and Ballala Sena's  Adbhuta Sagara   for their work  (Iyengar 2006).
 BS  had stated categorically that 
 it is not possible to determine by calculation the rising
 and setting of the comets. It had described in detail  the motion of a comet named Chala Ketu   (meaning, `moving comet')  underscoring its rise on the west and  increase in its size  as it moved  towards north,  touching   Ursa Major (Chandel \& Sharma 1991). 
  BS  had also delineated characteristics of 1000
 comets (Subbarayappa \& Sarma 1985).  It appears that it added 9
   to the existing number of  comets to make the number  a multiple of 10 (Miki
 and Yano, 2010). In India, the number nine was 
 auspicious probably because any region could be divided into  one
 central portion and  eight other  parts based on sub-division of regions along the four cardinal directions.

It is remarkable that, anticipating  periodic orbits, Narada had emphatically claimed - `there is only one comet which comes time and again', while  Bhadrabahu had reckoned that comets are   hundreds in number, each with different period (Sharma 1986, Chandel \& Sharma 1991). Some historians place the Jaina seer Bhadrabahu around 322 BC as a contemporary of Alexander the great,  and  as a preceptor of king Chandragupta Maurya  (Smith 1958). While, according to David Pingree (1983),  Bhadrabahu composed his samhita only about one or two centuries before al Biruni (973 - 1048 AD). 
Parashara ($\sim $ 1000 - 700 BC) had listed 101 comets, describing  features of 26 of them (including Chala Ketu discussed  earlier) which were likely to have been directly observed by him (Iyengar 2006). Morbid names like skull, bone, marrow, etc. were attached to some of the comets classified by Parashara  in the Death group of comets. As remarked earlier, even Atharvaveda associates comets with death since they look like the smoke rising from funeral pyre (Kochhar 2010).

 King of Mithila and Vanga, Ballal Sena ($\sim $ 1100-1200 AD), had compiled cometary records due to  seers Parashara, Vriddha Garga, Garga, Atharva, Varahamihira and  Asitadevala in his Adbhuta Sagara. Interestingly, it appears that while discussing the comet Dhuma Ketu, Vriddha Garga observes that it has a starry nature and that it ejects a jet of smoke in a  direction away from Sun before setting  (Iyengar 2006).

\section{4 Agastya, Nahusha, Saptarishi, and a possible comet}

Agastya  was a seer who composed about 27 hymns in the Rigveda (Mahadevan 1986).  He was referred to as Ugra or vigorous in this veda (Hiltebeitel 1977). It is curious to note that  vigorous or `ojas'  is cognate
with `aug'  (Gonda, 1972). 
Agastya has been  associated with the star Canopus or Alpha Argus since $\sim $  600 BC (Ghurye 1975). The  Indo-aryans could see 
 this bright star  only after reaching  latitudes lower than $37^\circ $.  Tradition has it that 
Agastya  was the first vedic seer to cross  Vindhyas to reach lower latitudes of the southern region of India. 
 The Rigvedic hymn CXCI (191) of book 1, most likely due to Mana's son Agastya, that  deplores  and complains about stings and bites from  poisonous aquatic worms, scorpions,  reptiles and nocturnal insects, states that it is the Sun who will provide  relief by scorching  and sucking up the venom, and then towards the end, mentions that the poison will be carried away by Krittika (Pleiades) like bearer girls transporting water in jars  (Griffith 1896).
 
There are three interesting points about this hymn. Firstly,  it claims that it is  the Sun drying up the water and thereby swallowing the poison that brings about a remedy, secondly it mentions Pleiades in connection with relief from venom, and  the third  is about water jars. Now, a later legend narrates that Agastya drained off an  entire ocean  to expose the dreadful demons hiding in the water. It is plausible that this myth grew out of  this hymn. 
The second point hints at the origin of the lore of seven sisters (Pleiades)  taking care of  newly born and protecting them from diseases. Thirdly,  Agastya  was said to  have been born in a jar, and was short in height. Was this hymn responsible for such a tale?  Or, being very tiny as a boy,  was he carried around  by bearer girls in a jar? In section 2, one had wondered whether the Proto-Dravidian word `kottu' (scorpion's sting) had led to the Sanskrit word `ketu' (comet). In the  southern regions of India, Agastya is also  associated with the Tamil word  `akatti' (Mahadevan 1986), which sounds similar not only to the seer's name but also to `kottu'. The above Rigvedic hymn also associates its composer to scorpion stings.  These are some interesting connections worth looking into.

The epic, Mahabharata, recounts the story of king Nahusha (known to  Rigveda, but with no myth  attached) who replaced  Indra as the king of the gods, as Indra went into hiding after slaying the demon Vrtra. After being anointed as king of the gods, Nahusha turned radiant with `five hundred lights on his forehead burning' as he absorbed energy from  gods, seers, demons, goblins, etc., and dominated the heavens (Hiltebeitel 1977). To impress Indra's consort Sachi, he ordered the seven seers (Ursa Major) to carry him in a palanquin. His arrogance infuriated the seer Bhrigu  who requested Agastya to become  a bearer of the  palanquin. Since, Agastya was short in height, the carriage tilted on one side when he substituted one of the seven seers. The resulting imbalance made  Nahusha very angry and he kicked Agastya, whereupon the latter cursed the former to turn into a serpent and fall from the heaven.

Hiltebeitel (1977) had linked the above tale with the origin of Deepawali (festival of lights) in India. However, the episode is more suggestive of Nahusha representing a comet that crossed the Big Bear from north and kept increasing in size as it moved southward towards Canopus (Agastya), and eventually disappearing beneath the horizon. According to the myths, Nahusha was a son of the daughter of Svarbhanu (later associated with Rahu and Ketu)  and belonged to the lunar dynasty with ancestors such as Atri (one of the stars of Ursa Major), Moon and  
Mercury (Hiltebeitel 1977). The hoary comet's overlap with the Big Bear could have conjured an imagery of its being carried by  Saptarishi  (it was  also referred  to as cart or  `wain' in the past; see Ghurye 1972, Hiltebeitel 1977). As the comet moved southwards, its tail got gradually elongated, finally making  an apparent contact with Canopus, which was depicted as Nahusha kicking Agastya. This is a reasonable conjecture. To buttress the surmise further, it is to be noted that  Varahamihira  prescribed worship of Canopus for kings, and warned that if this southern star  is struck by a
comet or a meteor there would be famine (verse 22 of BS; see Bhat 1981). Rigvedic hymn 191 of book 1 already describes Agastya's aversion towards reptiles, so one could imagine an antagonism between Canopus and  comets.  

 Furthermore, one may interpret Indra's hiding after Vrtra's assasination thereafter Nahusha getting anointed as  the lord of the heavens as  a symbolic portrayal of the supergiant star Antares or Jyeshtha (Indra?) getting  dimmer because of  its variability or disappearing due to lunar occultation,  while  the comet brightened and grew in size. Besides comets being linked with the serpentine Ketu (section 3),  Varahamihira, while referring to  Agastya's drinking up the ocean to reveal the hideouts of demons and  dangerous creatures, had likened the gems in the hoods of exposed snakes to comets (Kern  1870).
 
  It is not uncommon for a comet to become brighter as it moves from north to south.
For instance,  comet C/1853 G1 was discovered on April 5, 1853, south of $\rho $ aquilae (now in the northern constellation of Delphinus) by K. G. Schweizer. Then, it appeared in the southern hemisphere on April 30, 1853, with its tail pointing towards Canopus. It  grew from $\sim  4^\circ $ to $8^\circ - 10^\circ $   in extent within a day, and was last sighted on June 11, 1853 (Kronk 2003). It has an estimated period of about 782 years, and hence, could have been seen in $\simeq $ 493 BC and $\simeq $ 289 AD. There is also the case of a comet having a serpentine shape. The Chinese document of Se-ma Ts'ien had recorded that  
 the Standard of Tch'e-yeou  appeared  in 134 BC. It was comet-like but  arched backwards in the shape of a standard (Chavannes 1899).  According to Fotheringham (1919), this comet had also been identified by Hipparchus, as reported by the historian Pliny in Natural History. 
 
 Bruce Masse (2007) has given a very interesting interpretation of the episode of Vishnu's  Matsya Avatara (Fish incarnation) narrated in SBr, Mahabharata as well as  Puranas. In this mythology (very similar to the Biblical Noah's Ark tale),
 Manu, the progenitor of human race, had rescued a tiny but bright fish from a pond and had put it in a jar. The fish started growing fast so that it had to be transferred first to a river and then to the sea. It grew to become a gigantic horned fish in the ocean and warned Manu of an impending flood, and eventually saved him  by escorting Manu to a safer high altitude land. According to Masse,  Matsya Avatara represents an ancient comet which was small to begin with but grew as it neared the Sun and at the end it crashed into Indian ocean resulting in floods and tsunamis     
around May 10, 2807 BC. 

\section {5 Discussions}

In the distant past,  our ancestors made use of their terrestrial experiences along with large helpings of imagination to explain celestial phenomena. When they saw Sun or  Moon disappearing inch by inch during an eclipse, they pictured them being engulfed within the mouth of a  demon, in the way preys    get swallowed by pythons or  boas.  Darkness was an anathema to the `fire ritual' practicing  
priests, who monitored stellar positions to ascertain auspicious hours for vedic rites. Appearance of any bright heavenly object, in particular comets,  would have been  a
 cynosure of their eyes. Therefore, one speculates whether the hirsute nature of a comet's tail led them to conjure up  invisible aerial roots of a tree on the Pole star tying up the  stars and the planets to  keep them from falling on to Earth (section 2).
 
  Comets in India have been historically referred to as ketus and dhumaketus.  By the time of Varahamihira (6-th century AD), Ketu was depicted as the dismembered serpentine torso of the eclipse causing Rahu. Because of their tails and the term ketus for comets, it was natural  to associate  reptiles with comets.  It is also  worth exploring whether `ketu' is a Dravidian lone word, since a similar sounding  old Tamil word `kottu' is associated with scorpion's sting, pointed end or top tuft of hair (section 2).
   
  I have provided several arguments to support my surmise that the lore of the arrogant king Nahusha being carried by Saptarishi and ultimately thrown out of the heaven for kicking Agastya is an allusion to a bright comet of antiquity that crossed Ursa Major and moved southwards with its ever increasing tail size and went below the horizon after occulting Canopus. As an example, I cited the case of comet C/1853 G1 which was seen in the north near the star  $\rho $ aquilae in 1853, which then appeared in the southern sky with a long tail pointed at Canopus (section 4). If my conjecture is right, it would imply that the Nahusha episode was added to Mahabharata after comets got associated with serpentine forms.

Comets continue to influence human  imagination,  be it
 Giotto's fascination with Halley's comet leading him to depict
 it  as the star of Bethlehem in the 14th-century painting `The
 Adoration of the Magi' or the space scientists' determination to effect a soft landing of  Philae probe   on the nucleus of  comet 67P (section 1). We are living in happening times when some of the science fantasies that could be only dreamt of in the past, are materializing in front of our eyes. Some day, we may see exo-comets or even better, telescopes may detect tails behind  stars near the central blackhole of a quasar or a blazar due to the enormous radiation pressure from the accretion disc.   
 
  Endowed with memory and thinking ability, human beings have perceived and sought patterns in whatever  they could observe or experiment with. Periodic phenomena like  diurnal variations, apparent circling of stars around the Pole star, changes in lunar
 phases and the  cycle of seasons, etc. were all of paramount importance
 for human survival, explorations as well as  for practical conveniences. In order to keep track  of the functional aspects  of these natural effects, time markers and calendars came into being, which in turn with the  help of geometry and angle measurements, led to the development of positional astronomy. Thereafter, it was a natural jump  to Kepler's laws of planetary motion and then it got  catapulted to Newton's laws of gravitation,   encompassing not only cometary motions and stellar dynamics in galaxies but also  galactic dynamics in clusters and dark matter.

\section{acknowledgements}
 It is a pleasure to thank N. Rathnasree for her invaluable comments and Ujjwal Das Gupta for providing  technical support. 
\section{References}
\begin{itemize}
\item []Achar, B. N. N. 2000,  Indian Journal of History of
Science, 35, 173, and the references therein.
\item []Allen, R. H. 1963, Star names - their lore and meaning (Dover Publications Inc)
\item []Ansari, S. M. R. 1977, Bulletin of the Astronomical Society of India,  5, 10, and the references therein.
\item []Anthony, D. W. 2007, The Horse, the Wheel and Language: how bronze-age riders from the Eurasian Steppes shaped the modern world (New Jersey, Princeton University Press), and the references therein. 
\item []Bhat, R. M. 1981, Varahamihira’s Brhat Samhita (Delhi, Motilal Banarsidass Publishers Pvt. ltd.)
\item []Chandel, N. K. \& Sharma, S. 1991, Indian Journal of History of
 Science, 26, 375, and the references therein.
\item []Chattopadhyaya, D. 2008, in Cosmic Perspectives, edited by S. K. Biswas, D. C. V. Mallik \& C. V. Vishveshwara  (Cambridge, Cambridge University
 Press) pp.41-50, and the references therein.
\item []Chavannes, E. 1899, Les Memoires historiques de Se-ma Ts'ien, 3 
\item []Fotheringham, J. K. 1919, Monthly Notices of the Royal Astronomical Society, 79, 162, and the references therein.
\item []Ghurye, G. S. 1972, Two Brahmanical Institutions - Gotra and Charana (Bombay, Popular Prakashan) 
\item []Ghurye, G. S. 1977, Indian Acculturation: Agastya and Skanda  (Bombay, Popular Prakashan) 
\item []Gonda, J.  1972, The vedic god Mitra (Leiden)
\item []Griffith, Ralph T. H. (1896), The Hymns of the Rgveda  (Delhi, Reprinted by Motilal Banarasidass  Publishers Pvt. ltd., 1973)
\item []Hiltebeitel, A. 1977, History of Religions, 16, 329
\item []Iyengar, R. N. 2006, Journal of Geological Society of India, 67, 289
\item []Iyengar, R. N. 2010, Indian Journal of History of Science, 45,1
\item []Jacobi, H. G. 1909, Journal of the Royal Asiatic Society, 41, 721
\item []Kane, P. V. 1975,  History of Dharmasastra, Vol. 5 (Poona,  Bhandarkar Oriental Research Institute).
\item []Kern, H. 1870, Journal of the Royal Asiatic Society of Great Britain \and Ireland (New Series), 4,  430
\item[]Kronk, G. W. 2003, Cometography - A catalog of Comets, Vol.2 (Cambridge, Cambridge University Press), and the references therein.
\item []Kochhar, R.  2000, The Vedic People  (Delhi, Orient Longman)
\item []Kochhar, R. 2010, Indian Journal of History of Science, 45, 287, and the references therein.
\item []Mahadevan, I. 1986, Reprint from Journal of Tamil Studies, No.30 
\item []Markel, S. 1990, South Asian Studies, 6,  9
\item[]Masse, W. B. 2007, in Comet/asteroid impacts and human society  (Berlin, Springer,  Heidelberg)pp.25-70, and the references therein.
\item []Miki, M. \& Yano, M. 2010, Journal of Indian and Buddhist studies, 58, 1126
\item []Parpola, A. 2009, Scripta, 1, 37, and the references therein.
\item []Parpola, A. 2010, A Dravidian solution to the Indus script problem, a lecture delivered in the World Classical Tamil Conference, Coimbatore  (Chennai, Published by Central Institute of Classical Tamil)
\item []Pingree, D. E. 1981, Jyotihsastra: Astral and Mathematical Literature (Wiesbaden: Otto Harrassowitz), and the references therein.
\item []Pingree, D. 1983,  Journal of the  American Oriental Society 103, 353
\item []Saha, M. N. 1953, The Journal of the Royal Astronomical Society of
Canada, 47, 97 
\item []Sengupta, P. C. 1947, Ancient Indian Chronology (Calcutta, Calcutta University Press)
\item []Sharma, S. D.  1986, IAU Colloquium  on Oriental Astronomy (Oxford Press) pp.109-112.
\item []Smith, V. A. 1958, The  Oxford History of India  (Oxford) 3rd ed. 
\item []Stephenson, F. R.  1997, Historical Eclipses and Earth's Rotation (Cambridge, Cambridge University Press)
\item []Subbarayappa, B. V. \& Sarma, K. V. 1985, Indian Astronomy - A source-book (Bombay, Nehru Centre)
\item []Subbarayappa, B. V. 2008, in  Cosmic Perspectives, edited by  S. K. Biswas, D. C. V. Mallik \& C. V. Vishveshwara  (Cambridge, Cambridge University Press) pp.25-40.
\item []Vahia, M. N. \& Subbarayappa, B. V. 2011, Proceedings of the
Fourth Symposium on History of Astronomy, edited by   M. Soma \&
  K. Tanikawa (Japan, NAO) pp.16-19.
\item []Witzel, M. 1989, Dialectes dans les littératures indo-aryennes
\item []Witzel, M. 1995, in  Indian Philology and South Asian Studies. edited by George Erdosy, pp.85 - 125. 
\item []Witzel, M.  2003, in Blackwell Companion to Hinduism, edited 
by Gavin Flood (London, Blackwell Publishing Ltd.) pp.68 - 101, and the references therein.
\item []Whitney, W. D. 1905, Atharva-veda-samhita,  vol.2 ( Cambridge, Harvard University, USA)
\item []Yano, M. 2003, in Blackwell Companion to Hinduism, edited 
by Gavin Flood (London, Blackwell Publishing Ltd.) pp. 376 - 392, and the references therein.
\end{itemize}
\end{document}